\documentclass[%
 reprint,
 amsmath,amssymb,
 aps,
]{revtex4-2}

\usepackage{graphicx}
\usepackage{dcolumn}
\usepackage{bm}

\begin{document}

\title{Phase-sensitive x-ray ghost imaging}

\author{Margie P. Olbinado}
\email[]{margie.olbinado@esrf.fr}
\thanks{Corresponding author}
\affiliation{The European Synchrotron -- ESRF, CS40220, 38043 Grenoble, France}
\author{David M. Paganin}
\affiliation{School of Physics and Astronomy, Monash University, Victoria 3800, Australia}
\author{Yin Cheng}
\affiliation{The European Synchrotron -- ESRF, CS40220, 38043 Grenoble, France\, ~}
\author{Alexander Rack}
\affiliation{The European Synchrotron -- ESRF, CS40220, 38043 Grenoble, France\, ~}

\date{\today}

\begin{abstract}
Imaging with hard x-rays is an invaluable tool in medicine, biology, materials science, and cultural heritage. Propagation-based x-ray phase-contrast imaging \cite{Snigirev1995, Cloetens1996, Wilkins1996} and tomography have been mostly used to resolve micrometer-scale structures inside weakly absorbing objects as well as inside dense specimens. Indirect x-ray detection has been the key technology to achieve up to sub-micrometer spatial resolutions \cite{Koch1998}, albeit inefficiently and hence at the expense of increased radiation dose to the specimen. A promising approach to low-dose imaging and high spatial resolution even at high x-ray energies is ghost imaging \cite{Pelliccia2016, Yu2016, Schori2017, Pelliccia2018, Kingston2018, Zhang2018, Schori2018}, which could use single-pixel, yet efficient direct x-ray detectors made of high-density materials. However, phase contrast has not yet been realised with x-ray ghost imaging. We present an approach which exploits both the advantages of x-ray ghost imaging and the high sensitivity of phase-contrast imaging. In comparison with existing techniques, our method is efficient and achieves high-fidelity x-ray ghost images with phase contrast, accurate density resolution and dramatically higher spatial resolution. The method is scalable to practical tomography with large fields of view, micrometer spatial resolution, and with high-energy x-rays above 100 keV. It is also applicable to other phase-sensitive imaging techniques \cite{Momose2003, Pfeiffer2006, Morgan2012, Berujon2012} and with other probes such as neutrons, alpha rays, and muons, for which high spatial resolution detectors are limited or even not available.
\end{abstract}
\maketitle

\small\begin{figure*}
\includegraphics[width=180mm,angle=0]{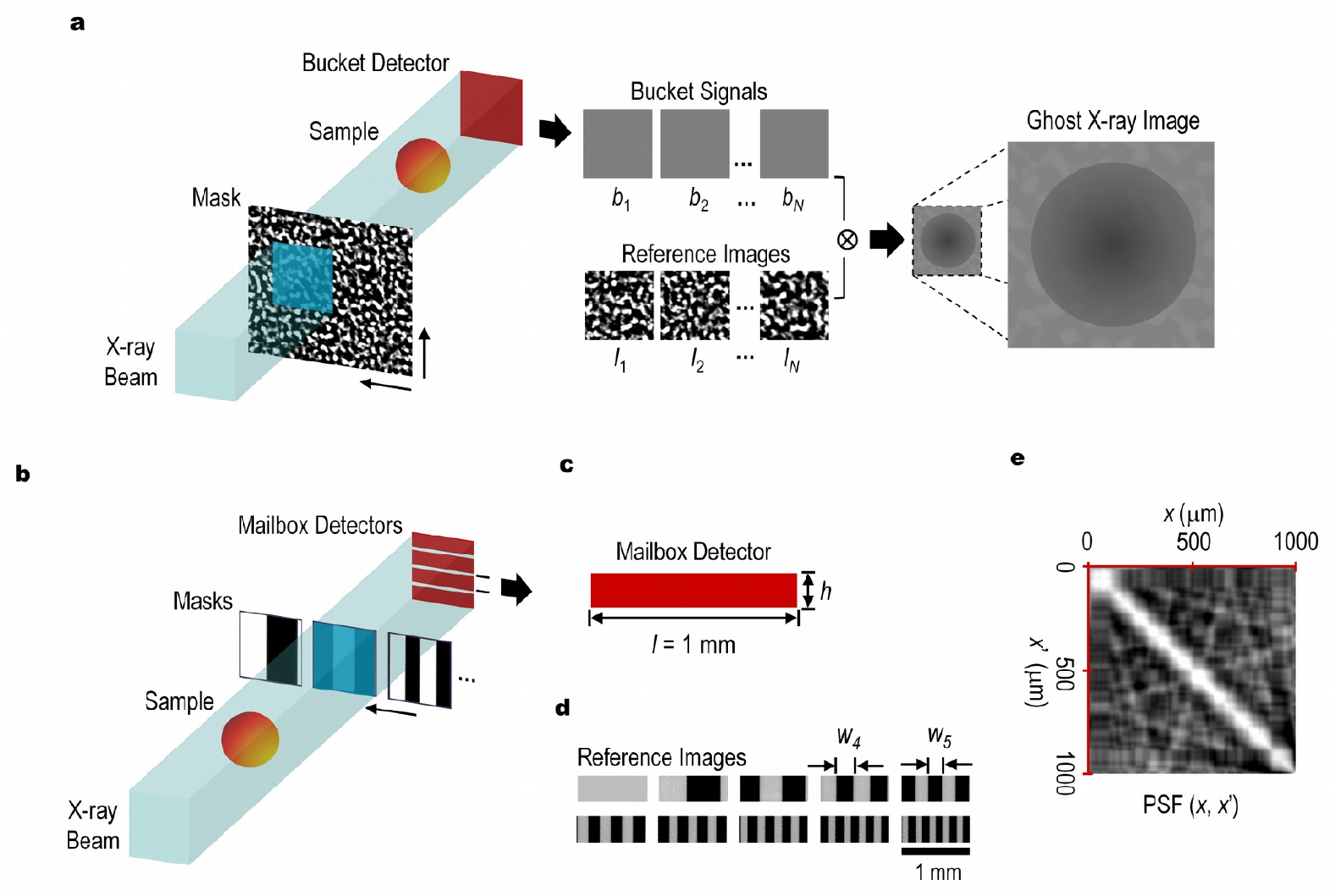}
\caption{\footnotesize\label{fig:setup} \textbf{Computational x-ray ghost imaging (XGI).} \textbf{a}, A conventional XGI set-up with a structured illumination approach using speckles. Intensity correlations between a series of known illuminating speckle fields (reference images) and the corresponding total intensity transmitted by the sample (bucket signals) collected by a bucket detector are utilised to synthesise an attenuation-contrast x-ray ghost image. \textbf{b}, Our phase-contrast XGI experimental set-up with a structured detection approach using periodic structures instead of speckles. By interchanging the mask and the sample in the sequence, the bucket signal is sensitive to x-ray phase shifts from the sample and a phase-contrast x-ray ghost image can be synthesised. 1D gratings are used as masks in combination with a 1D bucket detector, which is a collection of `mailbox' detectors. The ghost image reconstruction formula is applied for each mailbox detector (\textbf{c}). \textbf{d}, Reference x-ray images of $N$ = 10 grating patterns with 1 to 10 lines per mm. \textbf{e}, Calculated point-spread-function PSF$(x, x')$ of the phase-contrast x-ray ghost image without the sample for one of the mailbox detectors ($l$ = 1 mm, $N$ = 10). The spatial resolution (FWHM of the PSF) is $10\times$ the mailbox detector length.} 
\end{figure*}

X-ray ghost imaging \cite{Pelliccia2016, Yu2016, Schori2017, Pelliccia2018, Kingston2018, Zhang2018, Schori2018} is a newly developed imaging technique, derived from visible light optics \cite{Pittman1995, Strekalov1995, Katz2009, Bromberg2009}, which has the potential to achieve ultra-low radiation dose imaging and high spatial resolution. It utilises optical correlations between spatially resolved photons that never pass through an object of interest and non-spatially resolved photons that do pass through the object. Similarly to classical ghost imaging \cite{Katz2009, Bromberg2009}, experimental x-ray ghost imaging \cite{Pelliccia2016, Yu2016, Schori2017, Pelliccia2018, Kingston2018, Zhang2018} has been realised with speckled illumination. This has been done by using intrinsic noise of a synchrotron x-ray source \cite{Pelliccia2016}, and by phase-contrast-generated \cite{Yu2016, Schori2017, Pelliccia2018, Kingston2018} or attenuation-contrast-generated \cite{Zhang2018} speckle patterns. A ghost image is retrieved from intensity correlations between a series of speckle fields that illuminate an object and the total intensities transmitted by the object. Remarkably, the photons passing through the object are detected using only a so-called single-pixel `bucket' detector. Since single-pixel, high-Z (atomic number), direct x-ray detectors are significantly more efficient than two-dimensional (2D) indirect x-ray detectors commonly used in x-ray phase-contrast imaging, the potential reduction of the radiation dose to the specimens is highly appealing \cite{Pelliccia2016, Zhang2018}. The efficiency of indirect detectors is worse at high x-ray energies. Consequently, resolutions of only several micrometers have been achieved above 100 keV. In contrast, since the spatial resolution of a ghost image is determined by the speckle size and not by the detector \cite{Pelliccia2018, Ferri2010}, ghost imaging may achieve the much-coveted high spatial resolution with high x-ray energies. For example, large, single-pixel detectors made with the high-Z material CdTe have a quantum efficiency close to 100\% up to or even above 100 keV x-ray energy.

In essence, classical ghost imaging expresses the spatial distribution of an object's transmission function as a linear combination of speckle fields \cite{Katz2009, Bromberg2009}. Each speckle field is considered as a linearly independent basis vector. This is conceptually similar to building functions as a sum of sinusoids in a Fourier series. A classical ghost imaging setup is composed of a beam path (or reference arm) in which the speckle fields are detected by a 2D detector, and a beam path (or object arm) in which the object's transmitted intensities are detected by a bucket detector. In computational x-ray ghost imaging, the reference images are pre-determined or pre-characterised.

Figure~\ref{fig:setup}a shows a schematic diagram of a computational x-ray ghost imaging setup using speckle-generating masks, which are either x-ray absorbing or phase-shifting. A series of speckled illuminating fields is obtained by raster scanning the mask in the transverse plane.

Let $I_{\textup{\scriptsize in}}$ be a uniform, incident illumination intensity and $\hat{T}_{\textup{\scriptsize M}, \scriptsize j}(\textbf{r}_{\bot},z=R_{\textup{\scriptsize M}})$ be the transmission function of the \textit{j}th speckle mask (M) at a mask-to-sample propagation distance $R_{\textup{\scriptsize M}}$. Here, $\textbf{r}_{\bot} = (x,y)$ denotes coordinates in the planes perpendicular to the optical axis $z$. The $j$th illumination intensity onto the sample is 
\begin{equation}
I_{\scriptsize j}(\textbf{r}_{\bot})=\hat{T}_{\textup{\scriptsize M}, \scriptsize j}(\textbf{r}_{\bot},z=R_{\textup{\scriptsize M}})I_{\textup{\scriptsize in}}. 
\end{equation}
Letting $\hat{T}_{\textup{\scriptsize S}}(\textbf{r}_{\bot},z=R_{\textup{\scriptsize S}})$ be the transmission function of the sample (S) at the sample-to-bucket detector distance $R_{\textup{\scriptsize S}}$, the signal collected by the bucket detector may be written as: 
\begin{equation}
	b_{\scriptsize j} = \iint_\Omega\hat{T}_{\textup{\scriptsize S}}(\textbf{r}_{\bot},z=R_{\textup{\scriptsize S}})I_{\scriptsize j}(\textbf{r}_{\bot})\,d\textbf{r}_{\bot}
	\label{eq:bucket},
\end{equation}
where $\Omega$ is the surface over which the beam intensity is recorded and within which the sample is entirely contained. The hat (~$\hat{}$~) indicates that the transmission function is an operator and that the order of operation is crucial.  

Using $N$ speckle masks, an x-ray ghost image may be synthesised using \cite{Katz2009, Bromberg2009}: 
\begin{equation}
    G(\textbf{r}_{\bot})=\displaystyle\frac{1}{N}\displaystyle\sum^N_{\scriptsize j=1}(b_{\scriptsize j} - \bar{b})I_{\scriptsize j}(\textbf{r}_{\bot})
    \label{eq:ghost}.
\end{equation} 
The bucket signal $b_{\scriptsize j}$ subtracted by the mean $\bar{b}$ acts as a weighting coefficient for the corresponding reference speckle field $I_{\scriptsize j}$ in the superposition. Interestingly, even though neither of the detectors used in the measurement yields an image of the object, and though no photons passing through the object are ever registered by a position-sensitive detector, an image can be obtained by harnessing intensity {\em correlations}. 

This scheme, named classical ghost imaging with x-rays \cite{Pelliccia2016, Schori2017, Zhang2018, Pelliccia2018, Kingston2018} only retrieves the attenuation-contrast component of the object's x-ray transmission image independent of $R_{\textup{\scriptsize S}}$, thus losing the much-enhanced contrast or sensitivity to density variations in the sample that could be obtained by exploiting propagation-induced x-ray phase contrast. This inability to exploit phase contrast is the case with the usual mask--sample sequence. Here we show that the key to achieve phase contrast in x-ray ghost imaging is to reverse the sequence of sample and mask: a structured detection approach \cite{Duarte2008, Clemente2013, Zhang2015} instead of a structured illumination approach.

\begin{figure*}
\includegraphics[width=180mm,angle=0]{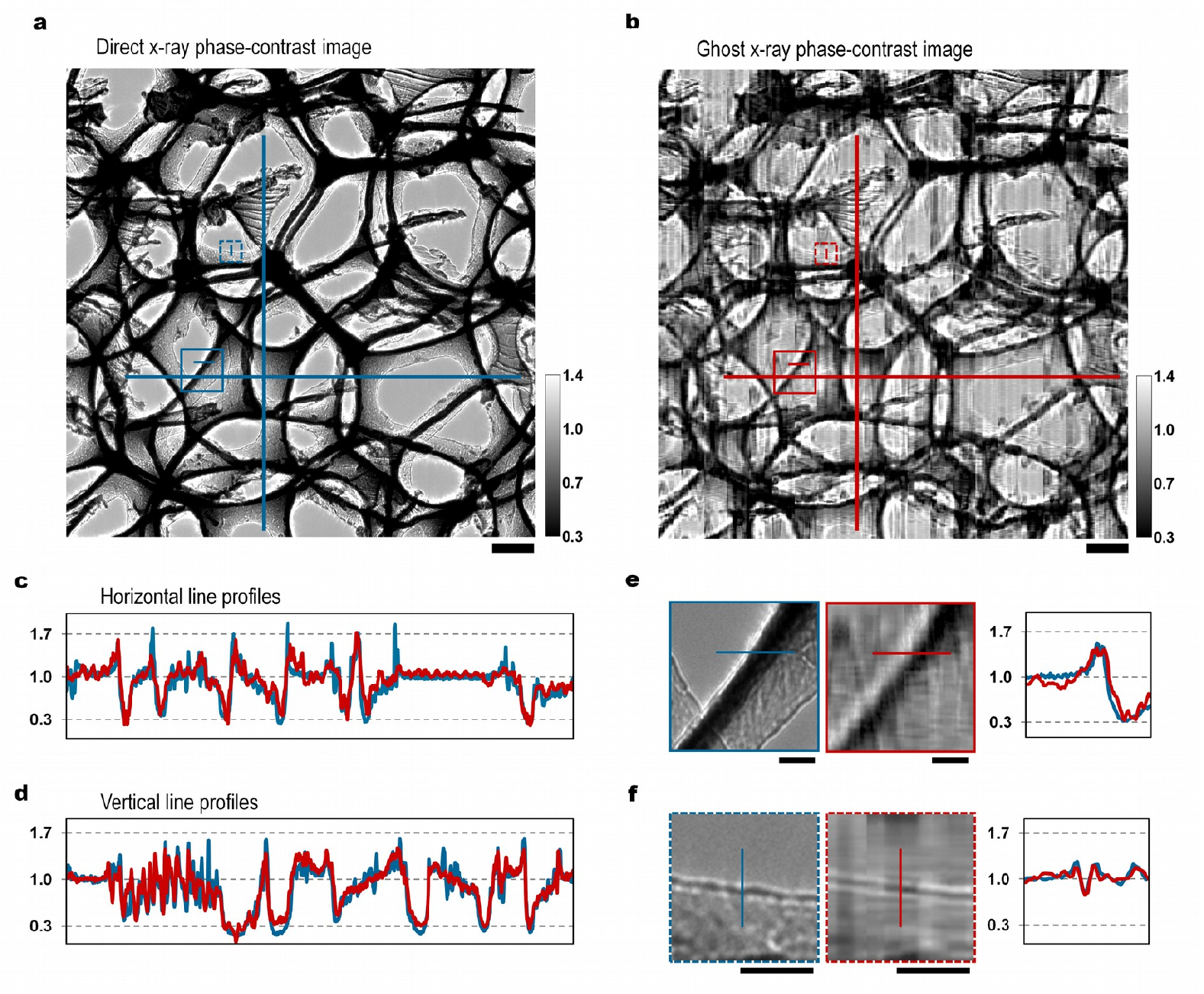}
\caption{\small\label{fig:result} \textbf{X-ray transmission images of interconnected aluminium lamellae cut from a sponge sample.} \textbf{a}, Direct x-ray phase-contrast image ($I_{\textup{\scriptsize S}}/I_{\textup{\scriptsize flat field}}$): calculated from radiographs directly recorded using a 2D imaging detector. \textbf{b}, Ghost x-ray phase-contrast image ($G_{\textup{\scriptsize S}}/G_{\textup{\scriptsize flat field}}$): calculated from synthesised ghost images both with the sample ($G_{\textup{\scriptsize S}}$) and without ($G_{\textup{\scriptsize flat field}}$). Scale bars in \textbf{a} and \textbf{b} represent 1 mm. \textbf{c}, Horizontal line profiles. \textbf{d}, Vertical line profiles. \textbf{e, f} Magnified views of the insets in \textbf{a} and \textbf{b} showing phase-contrast enhancement at the edges of representative thick and thin region of the lamellae. Scale bars in \textbf{e} and \textbf{f} represent 250 $\mu$m.}
\end{figure*}

The origin of contrast in an x-ray image is related to the object's complex refractive index: 
\begin{equation}
n(\textbf{r}_{\bot},z,\lambda)=1-\delta(\textbf{r}_{\bot},z,\lambda)+i\beta(\textbf{r}_{\bot},z,\lambda). 
\end{equation}
For simplicity, consider a homogeneous object of projected thickness $t(\textbf{r}_{\bot})$, and quasi-monochromatic, plane-wave x-ray radiation with wavelength $\lambda$ and initial intensity 
$I_{\textup{\scriptsize in}}$. Under the projection approximation, the spatial variations of the optical density, 
\begin{equation}
D(\textbf{r}_{\bot})=\frac{4\pi}{\lambda}\beta t(\textbf{r}_{\bot})
\end{equation}
and the x-ray phase shift, 
\begin{equation}
\phi(\textbf{r}_{\bot})=-\frac{2\pi}{\lambda}\delta t(\textbf{r}_{\bot})
\end{equation}
generate the attenuation contrast and phase contrast in the x-ray image, respectively \cite{Paganin2002}. In the hard x-ray regime, phase-contrast imaging is up to three orders of magnitude more sensitive than attenuation-contrast imaging. This is because absorption decreases with the fourth power of photon energy while phase contrast decreases with the square of the energy. 

A phase-contrast image is readily achieved through free-space propagation or Fresnel diffraction. At an object-to-detector propagation distance $R$ ($R \leq \frac{d^2}{\lambda}$, where $d$ is the characteristic length scale over which the object appreciably changes), a near-field phase-contrast image or Fresnel image may be derived from the so-called transport-of-intensity equation \cite{Teague1983} and is given by \cite{Paganin2002}:
\begin{eqnarray}
\hat{T}(\textbf{r}_\bot,z=R)~I_{\textup{\scriptsize in}}=\left(-\frac{R\delta}{\mu}\nabla_\bot^2+1\right)\textup{e}^{-\displaystyle\mu t(\textbf{r}_\bot)}I_{\textup{\scriptsize in}}
\label{eq:TIE},
\end{eqnarray}
where the linear attenuation coefficient $\mu=4\pi\beta/\lambda$ and $\nabla_\bot$ is the gradient operator in the $x$-$y$ plane. The transmission function contains both the the attenuation contrast and propagation-induced phase-contrast components. By conservation of energy, the average intensity of the Fresnel image over the entire area \textit{A} of the surface $\Omega(\textbf{r}_{\bot})$, at any distance $R$ is equal to that of the contact image ($R=0$).

By inserting Eqn.~\ref{eq:TIE} in Eqn.~\ref{eq:bucket} and invoking the conservation of energy, the resulting bucket signal is given by:
\begin{eqnarray}
b_{\scriptsize j} &=& \displaystyle\iint_\Omega\displaystyle \left(-\frac{R_\textup{\scriptsize S}\delta_\textup{\scriptsize S}}{\mu_\textup{\scriptsize S}}\nabla_\bot^2 + 1\right)\textup{e}^{-\displaystyle\mu_{\textup{\scriptsize S}} t_\textup{\scriptsize S}(\textbf{r}_\bot)}I_{\scriptsize j}(\textbf{r}_\bot)\,d\textbf{r}_\bot
\nonumber \\
&=& \displaystyle\iint_\Omega\displaystyle\textup {e}^{-\displaystyle\mu_{\textup{\scriptsize S}} t_{\textup{\scriptsize S}}(\textbf{r}_{\bot})}I_{\scriptsize j}(\textbf{r}_\bot)\,d\textbf{r}_\bot 
\label{eq:four}.
\end{eqnarray}
 Therefore, with the usual setup, where the mask is upstream of the sample, the bucket signal (Eqn.~\ref{eq:four}) is not sensitive to the propagation-induced phase-contrast component of $\hat{T}_{\textup{\scriptsize S}}$. Only the attenuation-contrast component $\exp[-\mu_{\textup{\tiny S}} t_{\textup{\tiny S}}(\textbf{r}_{\bot})]$ is recovered in the ghost image.
 
The key to achieving phase contrast in x-ray ghost imaging is to interchange the sequence of the sample and the mask in the beam. The mask is placed at the desired phase-contrast image plane, a distance \textit {R}$_{\textup{\scriptsize S}}$ downstream from the sample. The bucket signals, which measure the weighting coefficients of the superposition in the ghost image reconstruction (Eqn.~\ref{eq:ghost}), do not register a constant signal that is insensitive to phase contrast. We emphasise the crucial order of the sample and the mask by the order of operation of the transmission function in the bucket signal equation, which we write as:
\begin{eqnarray}
b_{\scriptsize j}=\displaystyle\iint_\Omega\hat{T}_{\textup{\scriptsize M}}(\textbf{r}_{\bot},z=R_{\textup{\scriptsize M}})\hat{T}_{\textup{\scriptsize S}}(\textbf{r}_{\bot},z=R_{\textup{\scriptsize S}})I_{\textup{\scriptsize in}}(\textbf{r}_{\bot})\,d\textbf{r}_{\bot}
\label{eq:five}.
\end{eqnarray}

In this way, the bucket signal is sensitive to the phase-contrast component of the sample's transmission image,  $\hat{T}_{\textup{\scriptsize S}}(\textbf{r}_{\bot},z=R_{\textup{\scriptsize S}})I_{\textup{\scriptsize in}}(\textbf{r}_{\bot})$. Notice also that depending on $R_{\textup{\scriptsize S}}$, a phase-contrast x-ray ghost image in the Fresnel (near-field) or Fraunhofer (far-field) regime can be synthesised. Furthermore, the ghost image synthesis can even be extended to any x-ray phase-contrast imaging approaches such as Talbot interferometry \cite{Momose2003, Pfeiffer2006}, and near-field speckle-tracking \cite{Morgan2012, Berujon2012}. This approach is also compatible with ghost imaging combined with x-ray diffraction topography and crystallography. Simulations have indeed shown that our detection approach is a means to achieve analyser-based x-ray phase-contrast ghost imaging \cite{Ceddia2018}. Equation \ref{eq:five} also clearly indicates that only the attenuation-contrast component of $\hat{T}_{\textup{\scriptsize M}}$ matters, hence amplitude masks should be used. 

Our setup for x-ray phase-contrast ghost imaging is depicted in Fig.~\ref{fig:setup}b. Instead of using speckle patterns, which form a non-orthogonal basis and require $N\gg p$ measurements in order to synthesise a ghost image consisting of $p$ pixels, we employ a linear combination of periodic fields with varying frequencies, which form a nearly-orthogonal set of basis patterns. This eliminates the inherent redundancy of the speckle-based approach. We implemented this using transmission gratings, which is practical since the fabrication of such gratings for hard x-rays is well-established. For example, high-aspect-ratio gratings with 2 $\mu$m pitch and 160 $\mu$m thickness used for up to 180 keV x-ray energy have been reported \cite{Ruiz-Yaniz2015}. Deterministic orthogonal basis patterns such as the Hadamard and Fourier basis patterns \cite{Duarte2008, Clemente2013, Zhang2015} have been used instead of non-orthogonal random patterns or speckles. Compressive sensing concepts \cite{Katz2009}, orthogonalisation of speckle fields \cite{Pelliccia2018} and iterative refinement \cite{Kingston2018} may also be employed to reduce $N$.

We opted not to use 2D gratings. Instead, we considered that a combination of 1D gratings with a 1D bucket detector is equivalent to using 2D gratings with a single-pixel (0D) detector. A 1D bucket detector constitutes a set of what we call mailbox detectors. Recently, high-Z, 2D  direct detectors such as the EIGER2 CdTe (Dectris Ltd., Switzerland) with pixel size of 75 $\mu$m have become commercially available. Line arrays or even pixels of such 2D detectors may be used as mailbox detectors. The ghost image reconstruction formula is applied for each mailbox detector (Fig.~\ref{fig:setup}c). The advantages are two-fold: (1) $N$ is reduced by a factor equal to the number of mailbox detectors used. (2) The spatial resolution may be tuned independently in two directions: one with the smallest grating line width $w$, and the other with the mailbox detector height $h$.

The position-dependent point-spread function (PSF) of an x-ray ghost image without sample was calculated for a given mailbox detector in order to check whether a set of gratings (Fig.~\ref{fig:setup}d) constitute a complete set of basis elements. The corresponding completeness relation \cite{Ferri2010, Pelliccia2018} that needs to be satisfied is given by:
\begin{equation}
\textup{PSF}(x-x')=\frac{1}{N}\sum^N_{\scriptsize j=1}[I_{\scriptsize j}(x') - \bar{I}][I_{\scriptsize j}(x) - \bar{I}],
\end{equation}
where $\bar{I}$ is the average intensity of the $j$th illumination instead of an average over all illuminations and $x$ runs over pixels equal to the mailbox detector length $l$. By way of example we show the calculated $\textup{PSF}(x-x')$ for $l$ = 1 mm and $N$ = 10 (Fig.~\ref{fig:setup}e) with a near-diagonal matrix, proving that the set of gratings is nearly orthogonal up to a resolution given by the width of the diagonal. The measured average full-width-at-half-maximum of the $\textup{PSF}(x-x')$, which represents the spatial resolution of the system, was 100 $\mu$m. This was expected and is equal to the smallest grating width $w_{N=10}$. With $N=p$, where $p = l/w_{10}$, the resulting periodic illuminating fields indeed constitute a nearly-orthogonal set. Note that neither orthogonalisation nor compressive sensing methods were applied prior to the calculation of the PSF.

Figure 2 shows a comparison of x-ray transmission images of interconnected aluminium lamellae cut from a sponge sample. The sample's transmission image ($I_{\textup{\scriptsize S}}/I_{\textup{\scriptsize flat field}}$) in Fig.~2a was calculated from images directly recorded by a 2D imaging detector. The sample's transmission image in Fig.~2b was calculated from synthesised ghost images both with the sample ($G_{\textup{\scriptsize S}}$) and without ($G_{\textup{\scriptsize flat field}}$). The phase-contrast-enhanced edges characteristic of a near-field Fresnel image \cite{Snigirev1995, Cloetens1996, Wilkins1996} are clearly visible in the ghost image just as in the direct image. The ghost image shows quantitative accuracy comparable to the direct image as illustrated in the line profiles shown in Figs.~2c-d. The transmission is unity at the pores, greater than unity at the material edges (phase contrast) and decreases at regions with increasing material density (attenuation contrast). Strong phase contrast, which is expected at a thick edge where a large x-ray phase gradient occurs, can be resolved along the horizontal (Fig.~2e). Due to the better spatial resolution along the vertical, a fringe pattern at the thin edge can be resolved (Fig.~2f). The white-black-white contrast is the familiar Fresnel diffraction phenomenon. The central minimum is due to destructive interference of waves symmetrical with the edge and the oscillations are Fresnel fringes \cite{Cloetens1996}. We emphasise that with attenuation contrast alone, this edge would have been invisible as the x-ray transmissions by the air and near the thin edge are essentially unity (see line profile in Fig.~2f). More importantly, this quantitative accuracy was achieved despite the fact that the resolution along the horizontal is an order of magnitude less than along the vertical ($w$ = 100 $\mu$m and $h$ = 6.5 $\mu$m). This ensures that the same high fidelity and quantitative accuracy can be achieved with large-pixel direct detectors when combined with micrometer-pitch gratings. For example, a combination of $w$ = 6.5 $\mu$m and $h$ = 100 $\mu$m should achieve a similar ghost image. The quantitative accuracy is a consequence of using a near-complete, near-orthogonal set of basis patterns versus a non-orthogonal set. Finally, we note that, similar to standard propagation-based phase-contrast imaging, phase-sensitive ghost imaging only requires sufficient spatial coherence via a small source size but essentially no temporal coherence. 

On the basis of these results we conclude that the presented method is a significant step forward in x-ray ghost imaging. The specific merit of our structured detection approach is the achievement of phase-contrast images with a resolution that goes beyond the spatial resolution of a bucket or mailbox detector. Furthermore, the use of gratings instead of speckles produces x-ray ghost images with high fidelity and quantitative accuracy. As the fabrication of large-area, high-aspect ratio transmission gratings is already well-established, the technique is easily scalable to large fields of view and micrometer spatial resolutions with high energy x-rays. Our method of x-ray phase-contrast ghost imaging can be implemented, and combined with phase retrieval and computed tomography. Our computational ghost imaging approach using gratings is also applicable with other probes such as neutrons, alpha rays, and muons, for which high spatial resolution detectors are limited or do not even exist. Single-pixel detectors would need to be combined with 2D gratings, or could be used with slits to form mailbox detectors that can be used with 1D gratings.

\subsection*{Methods}
The experiment was carried out at beamline ID19 of The European Synchrotron –- ESRF (Grenoble, France). A U-17 type undulator was used, with the gap tuned to generate 19 keV pink x-ray beam. The vertical and horizontal x-ray source sizes (full-width-at-half-maximum) were 25 $\mu$m and 150 $\mu$m, respectively. The sample was located 140 m from the source, while the mask and the detector were located 13 m from the sample. The sample was a metal foam (Mayser GmbH \& Co. KG, Germany) made of 99.7\% aluminium and with average pore size of 2.5 mm. An off-the-shelf x-ray test pattern (Type 23, H\"{u}ttner, Germany) composed of 0.5 mm thick lead patterns on 1 mm plexiglass was used for the gratings. The x-ray transmissions through the plexiglass and the lead were 96\% and 2\%, respectively. The 10 grating patterns used have 1 to 10 lines per mm. The set of 10 reference images (Fig.~1d) was obtained by scanning the test pattern in steps of 65 $\mu$m parallel with the middle grating line. This scanning was repeated for the measurements with the sample. An indirect x-ray image detector composed of an sCMOS camera (pco.edge; pixel size: 6.5 $\mu$m, PCO AG, Germany) coupled with a 100 $\mu$m-thick LuAG:Ce scintillator using a tandem of lenses (Hasselblad, Sweden) with 100 mm focal lengths was used. The mailbox detectors used were line arrays of 1000 $\mu$m length and 6.5 $\mu$m width (of the same indirect x-ray detector). A total of 1700 line arrays was used in the presented ghost image (Fig.~2b). The 11-mm horizontal field of view was achieved by stitching 21 images with 500 $\mu$m width each, that were cropped from 1 mm width ghost images. Both the scanning and the stitching would not have been necessary if we had a large rectangular grating.  

\small\bibliography{references}

\end{document}